# A digital interface for Gaussian relay and interference networks: Lifting codes from the discrete superposition model

M. Anand, *Student Member*, *IEEE*, and P. R. Kumar, *Fellow*, *IEEE*

*Abstract*—For every Gaussian network, there exists a corresponding deterministic network called the discrete superposition network [1]. We show that this discrete superposition network provides a near-optimal digital interface for operating a class consisting of many Gaussian networks in the sense that *any* code for the discrete superposition network can be naturally lifted to a corresponding code for the Gaussian network, while achieving a rate that is no more than a constant number of bits lesser than the rate it achieves for the discrete superposition network. This constant depends only on the number of nodes in the network and not on the channel gains or SNR. Moreover the capacities of the two networks are within a constant of each other, again independent of channel gains and SNR. We show that the class of Gaussian networks for which this interface property holds includes relay networks with a single source-destination pair, interference networks, multicast networks, and the counterparts of these networks with multiple transmit and receive antennas.

The code for the Gaussian relay network can be obtained from any code for the discrete superposition network simply by pruning it. This lifting scheme establishes that the superposition model can indeed potentially serve as a strong surrogate for designing codes for Gaussian relay networks.

We present similar results for the $K \times K$ Gaussian interference network, MIMO Gaussian interference networks, MIMO Gaussian relay networks, and multicast networks, with the constant gap depending additionally on the number of antennas in case of MIMO networks.

*Index Terms*—approximate capacity, deterministic model, discrete superposition model, interference networks, relay networks

## I. INTRODUCTION

Computing the capacities of wireless networks is a formidable problem. It has been a central theme of network information theory, where research has attempted to characterize the information-theoretic limits of data transmission in various abstract models of networks. Of potentially equal importance, and also of great interest, has been the problem of construction of optimal coding schemes which achieve the promised rates. However, except for some examples of networks like the MAC channel [33], [34] and the broadcast channel [31], [32], it has so far not been possible to accurately compute the capacities of even simple networks like the relay channel [14], [15] involving a small number of nodes.

M. Anand and P. R. Kumar are with the Dept. of ECE and Coordinated Science Lab, University of Illinois, Urbana, IL, USA. ({amurali2, prkumar}@illinois.edu.)

This material is based upon work partially supported by NSF under contracts CNS-0905397 and CCF-0939370, and AFOSR under Contract FA9550-09-0121.

This has motivated the pioneering line of work in [4]–[7] which aims at developing a class of noiseless deterministic networks that can serve as surrogates for Gaussian networks and suggest near-optimal coding schemes for Gaussian networks. Here the stringent goals of network information theory are relaxed by allowing for answers which are within a bounded gap of the capacity. By *bounded gap* is meant a constant that is independent of the channel gains or SNR, and is a function of only the number of nodes in the network. This relaxation in the problem has yielded new insights which have inspired construction of near-optimal coding schemes and improved our understanding of the fundamental limits imposed by various network configurations [3], [11], [8], [30], [10].

In this paper, we focus on rigorously studying the use of deterministic models both for approximating the capacity of relay and interference networks, as well as for the construction of coding schemes. We prove that a particular deterministic model, called the *discrete superposition model*, can serve as a digital interface for operating these Gaussian networks in that any coding scheme for this discrete superposition model can be naturally lifted to the Gaussian networks with no more than a bounded loss in the rate for each user.

### A. Relay networks

The relay channel was introduced by Van der Muelen in [14], and certain achievable rates were determined. Cover and El Gamal [15] subsequently studied the network in detail and developed two coding schemes which are now known as decode-and-forward and compress-and-forward. They also provided an upper bound on the capacity of the relay channel; which is now referred to as the cut-set bound. The cut-set bound was developed in full generality for networks in [16], [13].

The decode-and-forward scheme was extended to networks with many relays in [17], [18], and compress-and-forward was extended to larger networks in [19]. All these papers introduced certain novelties in existing coding schemes, though the best upper bound on the capacity of relay networks continues to be the cut-set bound. These schemes do not generally provide a guarantee of near optimality, in the sense that the gap between the achievable rates for any of these coding schemes and the cut-set bound could be arbitrarily high as a function of SNR. So, in effect, it has not been clear whether it is the cut-set bound that is weak or whether it is the coding schemes that are not exploiting all the features available in the wireless network.

This impasse motivated the work of [4], which studied coding schemes with provable guarantees on performance. This was done through the approach of introducing a *linear deterministic model*, where the linearity is over a finite field. The linear deterministic model captures the broadcast and interference aspects of wireless networks. The capacity of linear deterministic relay networks was determined to be precisely the same as the cut-set bound, which was additionally shown to be achievable by random coding at the nodes. Further, coding schemes were developed for general deterministic relay networks in [5], where the received signal at a node is a function of the signals transmitted by the neighboring nodes. The coding schemes for the linear deterministic model motivated the construction of coding schemes for Gaussian relay networks in [3], [6], [7], which are provably near-optimal for relay networks in the sense that they achieve all rates within a bounded gap from the cut-set bound. This result also establishes as a corollary that the cut-set bound for the Gaussian relay network is indeed approximately the capacity of the Gaussian relay network. Recently, the gap of between the achievable rate for the relay network and the cut-set bound has been further reduced by a generalization of compress-and-forward [9].

However, the above-mentioned results do not establish the closeness of the capacities of the linear deterministic network and the Gaussian network. Also, the near-optimal coding schemes for the Gaussian network are certainly inspired by the coding scheme for the linear deterministic counterpart, but there is no rigorous procedure to design a coding scheme for the Gaussian network given a coding scheme for the linear deterministic network, and consequently also no procedure for lifting codes from one to the other while preserving near-optimality. Hence it is unclear if a deterministic model only serves to aid the intuition in the construction of coding schemes for Gaussian networks, or if there is a more fundamental connection between the capacities of the deterministic network and the Gaussian network. More importantly, it is not clear if a coding strategy for a deterministic model can be explicitly used for designing a coding strategy for the Gaussian network that achieves comparable performance. We answer these questions in the affirmative here.

The linear deterministic model however does not approximate the capacity of Gaussian networks in general [3], [7], [1]. The question that therefore arises is whether there is a procedure for constructing for every Gaussian relay network a corresponding deterministic network in such a way that the capacities of the Gaussian and the deterministic network are within a bounded gap. This has been done in [3], [7] and [1] via different models. In the truncated deterministic model in [3], [7] the channel gains and inputs are complex valued, while the complex channel output has integer real and imaginary parts. The model used in [1], the discrete superposition model, is a somewhat more discrete model in the sense that channel gains and inputs are discrete valued. Hence corresponding to any Gaussian network, one can indeed construct a discrete superposition network which has the same topology as the Gaussian network and captures the effects of broadcast, interference, and attenuation by channel gains in a wireless network (see Section II-C). As mentioned in [3], [1], the bounded gap in capacities of the deterministic and Gaussian model does not necessarily imply a correspondence between coding schemes for the two models.

The next question that arises in this program of rigorous approximation of Gaussian networks by deterministic networks is whether one can also rigorously recover, with performance not decreasing by more than a constant number of bits, near-optimal coding schemes for Gaussian relay networks from coding schemes for discrete superposition relay networks. We establish this result in the affirmative. In fact, we show a stronger property that *every* coding scheme for the discrete superposition network can be simply mapped to a similar scheme for the Gaussian relay network in such a way that it continues to provide comparable performance. The lifting procedure therefore works uniformly over the class of all coding schemes, producing uniformly close performance. The lifting procedure is particularly natural and consists essentially of just pruning the codewords and using jointly typical decoding.

Thus the discrete superposition model provides a near-optimal digital interface for operating a Gaussian relay network. We extend this correspondence to MIMO relay networks and multicast networks. The superposition model may potentially be easier to design codes for than the Gaussian model since noise has been eliminated from the network and the sets of inputs and outputs are finite. Potentially, perhaps, wireless network coding could be useful in studying superposition networks, and subsequently Gaussian networks. This remains an open and intriguing question.

*B. Interference networks*

Interference networks have received much attention recently, following the results in [20]. In [20], the capacity of the interference channel with two transmitters and two receivers is determined within a constant gap of 1 bit. The near-optimal coding scheme in [20] is a specific choice among the myriad strategies proposed in [21]. A simpler proof of the result in [20] is provided in [24]. This was independently strengthened in [26], [27], and [28] where treating the interference as noise is shown to be capacity achieving for a restricted range of the parameters corresponding to a low interference regime. The capacity region of the $2 \times 2$ deterministic interference channel was determined in [22]. In [8], it is shown that the capacities of the linear deterministic interference channel and the Gaussian interference channel are within a bounded gap. A variant of the discrete superposition model was first used in [8] in a sequence of networks that reduced the $2 \times 2$ Gaussian interference channel to a linear deterministic interference channel.

Much less is known for Gaussian interference networks with more than two users. The number of degrees of freedom of the time-varying interference channel was characterized in [25] using the idea of interference alignment, and they were characterized for specific interference channels with fixed gains in [30]. Generalized degrees-of-freedom region of the fully symmetric many-user interference channel was computed in [29]. In general, the capacity region of the interference networks with three or more users is unknown, even to within a constant gap.

In this paper we prove that the rate region of the Gaussian interference network and the corresponding discrete superposition interference network are within a bounded number of bits. In fact, this correspondence extends to the case where the nodes have multiple transmit and receive antennas.

We can also prove the stronger result that the discrete superposition model provides a digital interface for operating the Gaussian interference network, in the sense, as above, that any coding scheme for the discrete superposition interference network can be lifted naturally to a scheme for the Gaussian interference network, with no more than a bounded loss in the rate for each user. This correspondence is extended to MIMO Gaussian interference networks where nodes can have multiple transmit and receive antennas.

*C. Outline of the paper*

In Section II, we introduce the various models used in this paper, viz. the Gaussian model, the linear deterministic model, and the discrete superposition model.

We develop the digital interface property for operating Gaussian relay networks using the discrete superposition model (Section III). The capacities of the Gaussian and discrete superposition relay networks are within a bounded gap. We rigorously prove a procedure for lifting any code for the discrete superposition relay network to the Gaussian relay network, and prove that there is no more than a bounded loss in the rate (Section III-B).

We show in Section IV that the rate regions of the Gaussian interference network and the discrete superposition interference networks are within a bounded gap. This proof subsumes the procedure for lifting a code from the discrete superposition interference network to the Gaussian interference network.

We also address the near-optimality of the digital interface defined by the discrete superposition model for MIMO channels, MIMO relay networks, MIMO interference networks, and multicast networks (Section V).

Notation: We will denote all random variables by lower case letters. Random vectors will be denoted by underbars, for example, as $\underline{x} = (x_1, x_2, \ldots, x_m, \ldots, x_N)$. If $\underline{x}$ is a complex vector, then the $m$-th term in the vector is given by $x_m = x_{mR} + \imath x_{mI}$, where $x_{mR}$ and $x_{mI}$ are the respective real and imaginary parts of $x_m$.

We denote the quantization of a complex number $x$ by $[x]$, where

$$[x] := \operatorname{sign}(x_R) \lfloor |x_R| \rfloor + \imath \operatorname{sign}(x_I) \lfloor |x_I| \rfloor. \quad (1)$$

In the above, $\lfloor \cdot \rfloor$ is the standard floor function.

Abusing notation, we will use the same letter to refer to the random variable and its realization. All logarithms in the paper are to the base 2.

## II. MODELS FOR NETWORKS

*A. Gaussian networks*

We begin by describing the class of Gaussian networks of interest. We consider a wireless network represented as a directed graph $(\mathcal{V}, \mathcal{E})$, where $\mathcal{V}$ represents the set of nodes, and the directed edges in $\mathcal{E}$ correspond to wireless links. In the case of relay networks, we label the nodes in $\mathcal{V}$ as $\{0, 1, \ldots, M\}$, where 0 is the source, $M$ is the destination, and the remaining are relay nodes. In the case of $K \times K$ interference networks, we divide the nodes in $\mathcal{V}$ into two sets; the first set $\{1, 2, \ldots, K\}$ consisting of the transmitters, and the second set $\{1, 2, \ldots, K\}$ consisting of the receivers. Though we use the same numbers to denote the sources and the destinations, it will be clear from the context which node we are referring to.

Denote by $h_{ij}$ the complex channel gain for link $(i, j) \in \mathcal{E}$. Let the complex number $x_i$ denote the transmission of node $i$. Every node has an average power constraint, taken to be 1. Node $j$ receives

$$y_j = \sum_{i \in \mathcal{N}(j)} h_{ij} x_i + z_j, \quad (2)$$

where $\mathcal{N}(j) = \{i : (i, j) \in \mathcal{E}\}$ is the set of its neighbors, $z_j$ is $\mathcal{CN}(0, 1)$ complex white Gaussian noise independent of transmitted signals, and $h_{ij} = h_{ijR} + \imath h_{ijI}$.

For MIMO networks, the nodes are allowed to have multiple transmit and receive antennas. The transmitted and received signals are described by vectors, where the number of elements in the vector corresponds to the number of transmit/receive antennas.

*B. The linear deterministic model*

The linear deterministic model was introduced in [4] as an approximate model to capture certain aspects of wireless networks. It should be noted that the linearity is with respect to the finite field $\mathbb{F}_2$.

The linear deterministic model is constructed based on the given Gaussian network as follows. We begin by choosing all the inputs and outputs of channels to be binary vectors of length $\max_{(i,j) \in \mathcal{E}} \lfloor \log |h_{ij}|^2 \rfloor$. Each link with channel gain $h$ in the Gaussian network is replaced by a matrix that shifts the input vector, allowing $\lfloor \log |h|^2 \rfloor$ most significant bits of the input to pass through. At a receiver, shifted binary vectors from multiple inputs are added bit by bit over the binary field. This models the partially destructive nature of interference in wireless. Modeling the broadcast feature of wireless networks, a node transmits the same vector on all outgoing links, albeit with different attenuations.

As an example, consider the Gaussian network in Fig. 1(a) and the corresponding linear deterministic network in Fig. 1(b). All inputs and outputs of channels are vectors in $\mathbb{F}_2$ of length 3. The channel is simply a linear transformation over the binary field. In the sequel, we will consider the high SNR scaling of a Gaussian network as all the channel gains $\{h_{ij} : (i, j) \in \mathcal{E}\}$ are scaled by a large positive constant $\gamma$ to give $\gamma h_{ij} : (i, j) \in \mathcal{E}$. We will study the relationship between the capacities of the high SNR scaled network and the linear deterministic approximation.

*1) Inability of the linear deterministic model to capture phase and power:* The linear deterministic model cannot capture the received signal power in certain Gaussian networks [3], [1]. Also, it does not capture the phase of the channel gain in a Gaussian network [1], as shown below via an example.

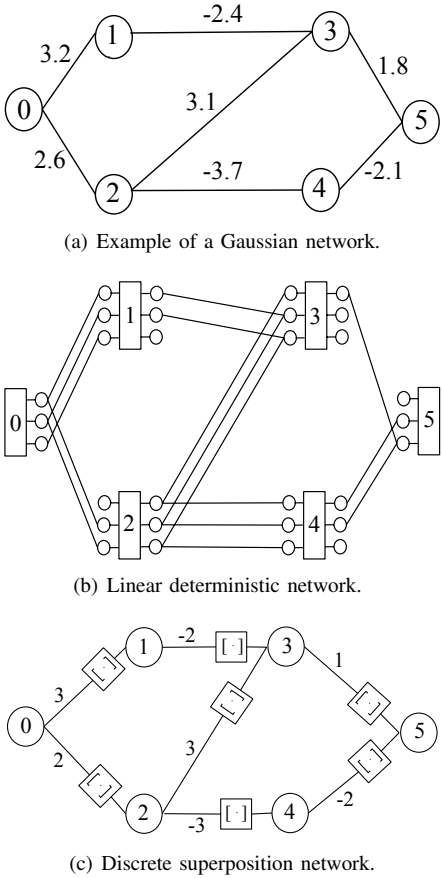

Fig. 1. A Gaussian network, linear deterministic counterpart, and discrete superposition network.

Due to these facts, in general, the model does not approximate the capacity of a Gaussian network to within a bounded number of bits.

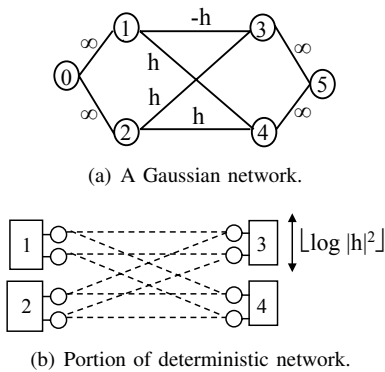

Fig. 2. Example showing that the linear deterministic model does not approximate the capacity of the Gaussian network to within a bounded number of bits.

*Example*: It is known that the cut-set bound is a near-optimal measure of the capacity of both the Gaussian and the linear deterministic network [3], [5]. Now consider the Gaussian network in Fig. 2 where the channels marked as $\infty$ have very high capacity. For this network, the cut-set bound is determined by the mutual information across the cut $\Omega = \{0,1,2\}$, which is given by[1]

$$I(x_\Omega; y_{\Omega^c}|x_{\Omega^c}) = \log|I + \mathcal{H}\mathcal{H}^\dagger|$$
$$= 4\log|h| + O(1), \text{ as } |h| \to \infty,$$

with

$$\mathcal{H} = \begin{bmatrix} -h & h \\ h & h \end{bmatrix}.$$

Hence, the capacity of the network is approximately $4\log|h|$.

A portion of the linear deterministic model for the Gaussian network is shown in Fig. 2(b). The capacity of the linear deterministic network is the rank of the transfer matrix of $\Omega = \{0,1,2\}$ which is $\lfloor 2\log|h| \rfloor$.

The gap between the capacities of the Gaussian network in Fig. 2(a) and its linear deterministic counterpart is therefore $2\log|h| + O(1)$. It is unbounded as $|h| \to \infty$.

The main reason for the unboundedness of the gap is that the linear deterministic model only considers the magnitude of a channel gain and effectively replaces each channel gain by a power of 2 that is nearest to its magnitude. Hence it does not capture the phase of the channel gain in Fig. 2(a). Constructing the deterministic model over a larger prime field (than $\mathbb{F}_2$) does not circumvent this problem.

### C. The discrete superposition model

The above example motivates the search for a deterministic model that does approximate the capacity of a Gaussian network to within a bounded number of bits. We now describe such a model. We associate a noiseless deterministic model with a Gaussian network, as follows. Let

$$n := \max_{(i,j) \in \mathcal{E}} \max\{\lfloor \log|h_{ijR}|\rfloor, \lfloor \log|h_{ijI}|\rfloor\}. \quad (3)$$

The inputs to the deterministic model are complex valued, with both real and imaginary parts taking values from the $2^n$ equally spaced discrete points $\{0, 2^{-n}, \ldots, 1 - 2^{-n}\}$, scaled by $1/\sqrt{2}$. The real or imaginary part of an input can accordingly be represented with the binary representation $x = \frac{1}{\sqrt{2}} \sum_{k=1}^{n} 2^{-k} x(k)$, with each $x(i) \in \mathbb{F}_2$.

Next we quantize the real and imaginary parts of channel gains in the Gaussian network to integers by neglecting their fractional parts, as in (1). The resulting quantized channel gain for link $(i,j)$ is given by $[h_{ij}]$.

The operation of the channel in the discrete superposition model is as follows. The channel between two nodes $i$ and $j$ in the discrete superposition network multiplies the input by the corresponding channel gain $[h_{ij}]$ and quantizes the product by neglecting the fractional components of both real and imaginary parts, i.e., it forms $[[h_{ij}]x_i]$. Note that $[[h_{ij}]x_i]$ lies in $\mathbb{Z} + \imath\mathbb{Z}$. These values are then added up at a receiver by the standard summation over $\mathbb{Z} + \imath\mathbb{Z}$. The received signal at node $j$ is therefore given by

$$y'_j = \sum_{i \in \mathcal{N}(j)} [[h_{ij}]x_i]. \quad (4)$$

---

[1] We chose i.i.d. $\mathcal{N}(0,1)$ inputs since this choice maximizes the mutual information, up to a constant [6].

Note that this model retains the essential *superposition* property of the wireless channel. Quantization of channel coefficients does not substantially change the channel matrix in the high SNR regime. Also, the blurring effect of noise is captured by constraining the inputs to positive fractions that can be represented by finite bits, as well as by quantization of the channel output.

Finally we note that the inputs in the discrete superposition model are restricted to be peak power constrained by the same value as the average power constraint on the transmit signals in the original Gaussian network. An important property of the discrete superposition model is therefore that the transmit signals in it are thus also valid for transmission in the Gaussian network. That is, encoder outputs in the discrete superposition network can also be used in the Gaussian network. We will see later in the sequel that this is important when we show how coding strategies in the discrete superposition network can be mapped to coding strategies in the Gaussian network.

For example, consider the Gaussian relay network in Fig. 1(a) and its discrete superposition counterpart in Fig. 1(c). The discrete superposition model preserves multiplication by channel gains and the superposition property of the channel, with the Gaussian noise replaced by quantization of outputs.

The discrete superposition model was first used in [8] in a sequence of networks that reduced the Gaussian interference channel to a linear deterministic interference channel. Here, it was shown that the capacities of the two-user Gaussian interference channel and the discrete superposition model of the same network are within a bounded gap. The model was generalized and given the name of a *superposition* model in [1] to distinguish it from the *linear* (over a vector space) deterministic model.

## III. A DIGITAL INTERFACE FOR GAUSSIAN RELAY NETWORKS

In this section, we study the class of Gaussian relay networks with a single source-destination pair. Here, the aim is to prove that the discrete superposition model for the relay network provides a near-optimal digital interface for operating the network. An example of a Gaussian relay network, with source 0, and destination 6 is shown in Fig. 1(a). In general, we consider networks with $M+1$ nodes, with the nodes labeled as $0, 1, \ldots, M$, where node 0 is the source, node $M$ is the destination, and the remaining nodes are relay nodes. The relay nodes have no information of their own to transmit, and assist the source in transmitting its data to the destination.

The description of a coding scheme for a relay network is somewhat involved due to the functions applied on the incoming data by the relay nodes. Also, due to the presence of cycles in the network, the encoding functions of the nodes can vary with time. In general, we define a $(2^{NR}, N)$ code for a relay network to be an encoding function for the source

$$\underline{x}_0 : \{1, 2, \ldots, 2^{NR}\} \to \mathcal{X}^N,$$

where $\mathcal{X}$ is the input alphabet of the channel, and a set of encoding functions for relay node $k$,

$$g_{k,m} : \mathcal{Y}_k^{m-1} \to \mathcal{X}, \text{ for } m = 1, 2, \ldots, N, k = 1, 2, \ldots, M-1,$$

where $\mathcal{Y}_k$ is the alphabet of node $k$'s received signal. For sake of simplicity, we assume that the input alphabet of each relay node is the same. As mentioned before, the encoding function of a relay node can vary with time, and the transmitted symbols can depend on all the received symbols previously received by the relay. The destination $M$'s decoding function is given by

$$g_M : \mathcal{Y}_M^N \to \{1, 2, \ldots, 2^{NR}\},$$

where $\mathcal{Y}_M$ is the alphabet of the received signal of the $M$-th node. Let $\mathcal{M}$ be a random variable uniformly distributed on $\{1, 2, \ldots, 2^{NR}\}$ that corresponds to the message that source 0 wants to communicate. Then $\mathcal{M}$ is mapped to the codeword $\underline{x}_0(\mathcal{M})$. The average probability of error is given by

$$P_e = Pr(g_M(\underline{y}_M) \neq \mathcal{M}),$$

where $\underline{y}_M$ is the signal received by the destination node $M$. The capacity of the relay channel is the supremum of all rates $R$ such that for any $\epsilon > 0$, there exists a blocklength $N$ for which $P_e < \epsilon$.

It is easy to show that any achievable rate below the capacity can be described by a multi-letter mutual information term:

*Lemma 3.1:* A rate $R$ lies in the capacity region of the relay network if and only if there exists a blocklength $N$, a distribution $p(\underline{x}_0)$, and a set of distributions $\{p(\underline{x}_k | \underline{y}_k), k = 1, 2, \ldots, M-1\}$ conforming to the causal constraints on the encoding functions at the relays as mentioned above such that

$$R < \frac{1}{N} I(\underline{x}_0; \underline{y}_M), \tag{5}$$

where $\underline{y}_M$ is the received signal at the destination.

*Proof:* We note a subtlety in computing the mutual information $I(\underline{x}_0; \underline{y}_M)$ where the effect of the various encoding functions applied at the relay nodes is captured in the joint distribution of $\underline{x}_0$ and $\underline{y}_M$.

Suppose we are given a distribution $p(\underline{x}_0)$ on input vectors of length $N$ and a set of relay encoding functions $\{p(\underline{x}_k | \underline{y}_k)\}$. Construct a collection of $2^{mNR}$ codewords for the source where each codeword is constructed by independently picking $m$ vectors, each of length $N$, with the distribution $p(\underline{x}_0)$ and appending them to get codewords of length $mN$. The encoding functions at the intermediate relay nodes are given by the transition probability kernels $\{p(\underline{x}_k | \underline{y}_k)\}$. The destination decodes by looking for a codeword jointly typical with its reception. Using standard random coding arguments (see [35]), the probability of error can be made arbitrarily small by considering large enough $m$. This proves the achievability of the rate.

For proving the converse, fix a $(2^{NR}, N)$ code and observe:

$$\begin{aligned} NR &= H(\mathcal{M}) \\ &= I(\mathcal{M}; \underline{y}_M) + H(\underline{y}_M | \mathcal{M}) \\ &\leq I(\mathcal{M}; \underline{y}_M) + 1 + P_e NR \tag{6} \\ &\leq I(\underline{x}_0; \underline{y}_M) + 1 + P_e NR, \tag{7} \end{aligned}$$

where we used Fano's lemma in (6) and the data processing inequality in (7). Since the rate $R$ is achievable, $P_e$ can be made arbitrarily small for sufficiently large $N$. On dividing both sides of (7) by $N$, we get the converse. ∎

The subtlety in the above proof regarding the mappings at the encoders makes the proof of the near-optimality of using the discrete superposition model as a digital interface for relay networks more involved than in the case of the interference channel, which is analyzed subsequently in a later section.

The next theorem indicates the relevance of the discrete superposition model in analyzing relay networks.

*Theorem 3.2:* Consider the relay network described above. The capacity $C_G$ of the Gaussian relay network and the capacity $C_D$ of the discrete superposition relay network is within a bounded gap of $\kappa_\mathcal{R}$ bits where

$$\kappa_\mathcal{R} = O(M \log M). \tag{8}$$

Note that the above theorem does *not* establish any correspondence between coding schemes for the two networks. We prove the converse part of the theorem next, i.e., $C_D \geq C_G - O(M \log M)$, while the remainder of the proof is presented in the subsequent section via a stronger argument concerning the lifting of coding schemes from the discrete superposition model to the Gaussian model.

### A. Capacity(Discrete superposition network) ≥ Capacity(Gaussian network), to within an additive constant

*Lemma 3.3:* Let $C_G$ be the capacity of the Gaussian relay network, and $C_D$ be the capacity of the discrete superposition relay network. Then,

$$C_D \geq C_G - O(M \log M). \tag{9}$$

*Proof:* The cut-set bound [16] on the capacity in a single source-destination pair relay network, with source 0 and destination $M$, is given by

$$C \leq \max_{p(x_0, x_1, \ldots, x_{M-1})} \min_{\Omega \in \Lambda} I(x_\Omega; y_{\Omega^c} | x_{\Omega^c}), \tag{10}$$

where $C$ is capacity of the network and $\Lambda$ is the set of all partitions of $\mathcal{V}$ with $0 \in \Omega$ and $M \in \Omega^c$. In [7] it is proved that the cut-set bound for a Gaussian network is achievable up to a bounded gap, where the gap is at most $O(M \log M)$. It is also proved that while computing the cut-set bound for a Gaussian relay network, we can choose the random variables in the optimization to be i.i.d. complex Gaussian $\mathcal{CN}(0, 1)$. With this choice, the cut-set bound evaluates to within $O(M \log M)$ of the maximum value.

It is also proved in [7] that the cut-set bound is achievable for a general class of deterministic networks, of which the discrete superposition network is a special case, when the maximization in the cut-set bound is restricted to independent random variables.

We consider a particular cut in a network. We start with the Gaussian model and choose the random variables corresponding to the cut to be i.i.d. Gaussian. We reduce the Gaussian network to a discrete superposition network in stages, bounding the loss in mutual information across the cut at each stage. At the end, we will have a distribution for the inputs for the discrete superposition network where all the inputs are independent. We will prove that the total loss in the mutual information as a result of these transformations is at most $O(M \log M)$. Repeating this across all the cuts in the network proves that the capacity of the discrete superposition network is within $O(M \log M)$ bits of the capacity of the Gaussian network.

Now consider a cut $\Omega$ containing $P$ nodes (including the source) and $\Omega^c$ containing $Q$ nodes (including the destination), where $P + Q = M + 1$. For the sake of simplicity, we assume that all the nodes in $\Omega$ are connected to all the nodes in $\Omega^c$, and the nodes in $\Omega^c$ are not connected among themselves[2]. The received signals in $\Omega^c$ are given by

$$y_j = \sum_{i=0}^{P-1} h_{ij} x_i + z_j, \; j = P, \ldots, M. \tag{11}$$

Choose $x_i$ as i.i.d. $\mathcal{CN}(0, 1)$. Each transmitted signal $x_i$ can be split into its quantized part $[x_i]$ and fractional part $\tilde{x}_i$, where

$$\tilde{x}_i := x_i - [x_i].$$

We discard the quantized part $[x_i]$ of all the transmitted signals and and retain $\tilde{x}_i$. Since $x_i$'s satisfies a unit average power constraint, $\tilde{x}_i$'s also satisfies a unit average power constraint. Define

$$\tilde{y}_j := \sum_{i=0}^{P-1} h_{ij} \tilde{x}_i + z_j, \; j = P, \ldots, M. \tag{12}$$

The discarded portion of the received signals is given by

$$\hat{y}_j := \sum_{i=0}^{P-1} h_{ij} [x_i], \; j = P, \ldots, M. \tag{13}$$

The mutual information across the cut $\Omega$ for channels (11) and (12) can be compared as

$$\begin{aligned}
I(x_\Omega; y_{\Omega^c}) &\leq I(x_\Omega; \tilde{y}_{\Omega^c}, \hat{y}_{\Omega^c}) \\
&\leq I(\tilde{x}_\Omega, [x_\Omega]; \tilde{y}_{\Omega^c}, \hat{y}_{\Omega^c}) \\
&= I(\tilde{x}_\Omega, [x_\Omega]; \tilde{y}_{\Omega^c}) + I(\tilde{x}_\Omega, [x_\Omega]; \hat{y}_{\Omega^c} | \tilde{y}_{\Omega^c}) \\
&= I(\tilde{x}_\Omega; \tilde{y}_{\Omega^c}) + I(\tilde{x}_\Omega, [x_\Omega]; \hat{y}_{\Omega^c} | \tilde{y}_{\Omega^c}) \quad (14) \\
&\leq I(\tilde{x}_\Omega; \tilde{y}_{\Omega^c}) + H(\hat{y}_{\Omega^c}) \\
&\leq I(\tilde{x}_\Omega; \tilde{y}_{\Omega^c}) + \sum_{i=0}^{P-1} H([x_i]), \tag{15}
\end{aligned}$$

where (14) follows because $[x_\Omega] \to \tilde{x}_\Omega \to \tilde{y}_{\Omega^c}$ form a Markov chain, and (15) holds because $\hat{y}_j$ is a function of $\{[x_i]\}$ from (13). It is proved in the appendix that $H([x_i]) \leq 6$; hence we get

$$I(\tilde{x}_\Omega; \tilde{y}_{\Omega^c}) \geq I(x_\Omega; y_{\Omega^c}) - 6P. \tag{16}$$

Since $\tilde{x}_{iR}$ and $\tilde{x}_{iI}$ lie in $(-1, 1)$, we obtain positive inputs by adding 1 to each. This is equivalent to adding $\sum_i h_{ij}(1+\imath)$

---

[2]This ensures that $I(x_\Omega; y_{\Omega^c} | x_{\Omega^c}) = I(x_\Omega; y_{\Omega^c})$. The more general case can also be similarly handled.

to $\tilde{y}_j$. We also divide by $2\sqrt{2}$ throughout to get:

$$(\tilde{y}_j + \sum_{i=0}^{P-1} h_{ij}(1+\imath))/2\sqrt{2}$$
$$= \sum_{i=0}^{P-1} h_{ij}(\tilde{x}_i + 1 + \imath)/2\sqrt{2} + z_j/2\sqrt{2}. \quad (17)$$

Lets denote the vector of $(1+\imath)$'s as $\nu_\Omega$ and note that

$$I(\tilde{x}_\Omega; \tilde{y}_{\Omega^c})$$
$$= I\left(\tilde{x}_\Omega, \frac{\tilde{x}_\Omega + \nu_\Omega}{2\sqrt{2}}; \tilde{y}_{\Omega^c}\right)$$
$$= I\left(\tilde{x}_\Omega, \frac{\tilde{x}_\Omega + \nu_\Omega}{2\sqrt{2}}; \tilde{y}_{\Omega^c}, \frac{\tilde{y}_{\Omega^c} + (\sum_i h_{ij}(1+\imath))_j}{2\sqrt{2}}\right)$$
$$= I\left(\frac{\tilde{x}_\Omega + \nu_\Omega}{2\sqrt{2}}; \frac{\tilde{y}_{\Omega^c} + (\sum_i h_{ij}(1+\imath))_j}{2\sqrt{2}}\right).$$

To avoid introducing new notation, for the rest of the proof we abuse notation and denote the left hand side of (17) by $y_j$, $(\tilde{x}_i + (1+\imath))/2\sqrt{2}$ by $x_i$, and $z_j/2\sqrt{2}$ by $z_j$, for all $j$.

With the new notation, $|x_i| \leq 1$, with positive real and imaginary parts, and $z_j$ is distributed as $\mathcal{CN}(0, 1/8)$.

The features of the model that we next address are:
1) channel gains are quantized to lie in $\mathbb{Z} + \imath\mathbb{Z}$,
2) real and imaginary parts of the scaled inputs are restricted to

$$n := \max_{(i,j)\in\mathcal{E}} \max\{\lfloor\log|h_{ijR}|\rfloor, \lfloor\log|h_{ijI}|\rfloor\} \quad (18)$$

bits,
3) there is no AWGN, and
4) outputs are quantized to lie in $\mathbb{Z} + \imath\mathbb{Z}$.

Let the binary expansion of $\sqrt{2}\, x_{iR}$ be $0.x_{iR}(1)x_{iR}(2)\ldots$, i.e.,

$$x_{iR} =: \frac{1}{\sqrt{2}} \sum_{p=1}^{\infty} 2^{-p} x_{iR}(p). \quad (19)$$

The received signal in the discrete superposition channel only retains the following relevant portion of the input signals:

$$y'_j = \sum_{i=0}^{P-1} [[h_{ij}]\, x'_i],\ j = 0,\ldots,P-1, \quad (20)$$

where

$$x'_{iR} := \frac{1}{\sqrt{2}} \sum_{p=1}^{n} x_{iR}(p) 2^{-p},$$
$$x'_{iI} := \frac{1}{\sqrt{2}} \sum_{p=1}^{n} x_{iI}(p) 2^{-p},$$

with $n$ defined in (3). To obtain (20) we subtracted $\delta_j$ from $y_j$, where

$$\delta_j := \sum_{i=0}^{P-1} \Big(h_{ij}(x_i - x'_i) + (h_{ij} - [h_{ij}])\, x'_i + \quad (21)$$
$$([h_{ij}]\, x'_i - [[h_{ij}]\, x'_i])\Big) + z_j$$
$$=: \sum_{i=0}^{P-1} w_{ij} + z_j$$
$$=: v_j + z_j.$$

To bound the loss in the mutual information across the cut in the discrete superposition network from the original Gaussian interference network, we have

$$I(x_\Omega; y_{\Omega^c})$$
$$\leq I(x_\Omega; y_{\Omega^c}, y'_{\Omega^c}, \delta_{\Omega^c})$$
$$= I(x_\Omega; y'_{\Omega^c}, \delta_{\Omega^c})$$
$$= I(x_\Omega; y'_{\Omega^c}) + I(x_\Omega; \delta_{\Omega^c}|y'_{\Omega^c})$$
$$= I(x_\Omega; y'_{\Omega^c}) + h(\delta_{\Omega^c}|y'_{\Omega^c}) - h(\delta_{\Omega^c}|y'_{\Omega^c}, x_\Omega)$$
$$\leq I(x'_\Omega; y'_{\Omega^c}) + h(\delta_{\Omega^c}) - h(\delta_{\Omega^c}|y'_{\Omega^c}, x_\Omega, \nu_\Omega)$$
$$\leq I(x'_\Omega; y'_{\Omega^c}) + \sum_{j=P}^{M}(h(\delta_j) - h(z_j))$$
$$= I(x'_\Omega; y'_{\Omega^c}) + \sum_{j=P}^{M} I(v_j; \delta_j).$$

By bounding the magnitudes of the terms in (21), we get $|w_{ij}| \leq 3\sqrt{2}$. So, $I(v_j; \delta_j)$ is at most the mutual information of a Gaussian MISO channel with average input power constraint less than $(3\sqrt{2})^2 P \leq 18P$ and

$$I(v_j; \delta_j) \leq \log(1 + 18P/(1/8))$$
$$< \log(1 + 144P). \quad (22)$$

Hence we get

$$I(x'_\Omega; y'_{\Omega^c}) \geq I(x_\Omega; y_{\Omega^c}) - Q\log(1 + 144P). \quad (23)$$

Note that $I(x'_j; y'_j)$ is the mutual information across the cut $\Omega$ in the discrete superposition network. By accumulating the losses in transforming the inputs for the Gaussian channel into the corresponding inputs for the deterministic channel in (16) and (23), we obtain

$$I(x'_\Omega; y'_{\Omega^c}) \geq I(x_\Omega; y_{\Omega^c}) - O(M \log M), \quad (24)$$

where $x_\Omega$ and $y_{\Omega^c}$ in the above equation are the respective channel inputs and outputs in the original Gaussian cut. Also, we began by choosing the inputs $x_i$ in the Gaussian network to be i.i.d. Gaussian $\mathcal{CN}(0,1)$. At the end of the transformations, the channel inputs in the discrete superposition network $x'_i$ are also independent. This completes the proof. ∎

*B. Lifting a coding scheme for the discrete superposition network to the Gaussian network*

A coding strategy for either the Gaussian or superposition relay network specifies codewords transmitted by the source, a mapping from the received signal to a transmit signal for every relay node, and a decoding function for the destination. For sake of simplicity of exposition, we assume that the graph describing the network is acyclic and that the relays employ time-invariant encoding functions. Later, in Section. III-D6, we mention how to handle more general encoding functions.

We describe how to lift a coding strategy for the discrete superposition network to a strategy for the Gaussian network.

Consider a $(2^{NR}, N)$ code for the discrete superposition network with zero probability of error, for a certain $N$. The probability of error can be reduced to zero due to the deterministic nature of the network; see Sec. III-D1.

Denote the block of $N$ transmissions at node $j$ in the discrete superposition network by an $N$-dimensional transmit vector $\underline{x}_j$, and similarly the received vector by $\underline{y}'_j$. All signals in the discrete superposition network are a (deterministic) function of the codeword $\underline{x}_0$ transmitted by the source.

Next, we build a $(2^{mNR}, mN)$ code, denoted by $\mathcal{C}_0$, for the discrete superposition network, with every $mN$-length codeword constructed by adjoining $m$ codewords from the old code, for a large $m$. This is again a rate $R$ code since it simply uses the old code $m$ times on the superposition network. We can visualize the construction of codewords in $\mathcal{C}_0$ by referring to Fig. 3.

In the $(2^{mNR}, mN)$ code, node $j$
1) breaks up its received signal, denoted by $\underline{\mathbf{y}}'_j$, into $m$ blocks of length $N$,
2) applies the mapping used in the $(2^{NR}, N)$ code on each of the $m$ blocks to generate $m$ blocks of transmit signals,
3) and adjoins $m$ blocks of transmit signals to construct a new transmit signal, denoted by $\underline{\mathbf{x}}_j$, of length $mN$.

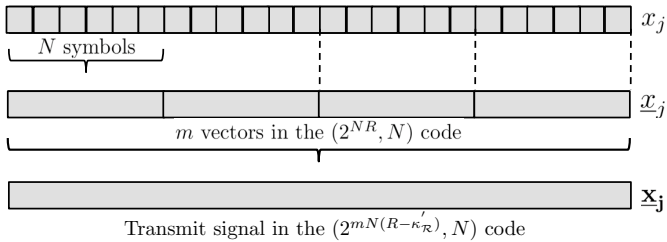

Fig. 3. Relationship among the signals transmitted by node $j$.

As shown in Fig. 3, the relationship between various signals associated with the transmission of node $j$ is akin to packetization in computer networks.

A subset of the codewords of $\mathcal{C}_0$, defined below, forms the set of codewords of the code for the Gaussian relay network.

Pruning the set of codewords: Node $j$ has a finite set of $\epsilon$-strongly typical $\underline{\mathbf{y}}'_j$'s (see [35]) in the code for the superposition network. We randomly, i.e., independently and uniformly, pick a $2^{-m(N\kappa+2\eta)}$ fraction of them and denote the resulting set by $\mathcal{S}_j$. $\kappa > 0$ is defined later in (30) as a function only of the number of nodes in the network and not the channel gains, while $\eta > 0$ is specified later and can be made arbitrarily small. We repeat this pruning procedure for all the nodes.

Denote the intersection of the inverse images of $\mathcal{S}_j$ in $\mathcal{C}_0$, for $j = 1, 2, \cdots, M$, by $\mathcal{C}_G$. Transmission of any vector in $\mathcal{C}_G$ results in the received vector at node $j$ belonging to $\mathcal{S}_j$ in the discrete superposition network. $\mathcal{C}_G$ forms the set of codewords for the Gaussian network.

Encoding and decoding procedure in the Gaussian network: The source in the Gaussian network transmits a codeword $\underline{\mathbf{x}}_0$ from $\mathcal{C}_G$. Assume throughout that node 1 can listen, i.e., has a link, to the source. Node 1 receives a noisy signal and decodes to a vector in $\mathcal{S}_1$. We will specify in the sequel how this decoding is to be done. Then, using the encoding function from $\mathcal{C}_0$, it constructs its transmit signal. All relay nodes operate in a similar way. Finally, the destination decodes its noisy reception to a signal in $\mathcal{S}_M$, and maps it to a codeword by simply using the decoding function from $\mathcal{C}_0$.

Note that we are operating the Gaussian network over the digital interface naturally defined by the signals transmitted and received in the corresponding discrete superposition network.

We summarize the main result concerning the lifting procedure in Theorem 3.4.

*Theorem 3.4:* Consider a Gaussian network with a single source-destination pair and $M - 1$ relay nodes, and consider a code for the discrete superposition model of the network that communicates at a rate $R$.

*Then, the lifting procedure and the digital interface defined by the discrete superposition model yield a code for the original Gaussian network that communicates at a rate $R - \kappa'_\mathcal{R}$, where*

$$\kappa'_\mathcal{R} := M(\log(6M - 1) + 10). \quad (25)$$

It should be noted that $\kappa'_\mathcal{R}$ does not depend on the channel gains. Therefore, the above theorem provides a lifting procedure that attains a rate in the Gaussian network within a bounded amount of $R$ at any SNR.

The theorem also proves that the capacity of the Gaussian network is at least as large as the capacity of the discrete superposition network, up to a bounded gap $\kappa'_\mathcal{R}$. This completes the proof of Theorem 3.2. Also, the above theorem applies to any coding scheme for the superposition network and, in particular, to an optimal scheme. Since the capacities of the Gaussian and the superposition network are within a bounded gap, the optimal scheme for the superposition network can be lifted to obtain a near-optimal coding scheme for the Gaussian network.

### C. A genie-based argument

Before delving into the details of the proof, we start with a genie-based argument explaining the ideas behind the proof of Theorem 3.4. The theorem is subsequently proved in detail in Sec. III-D.

The arguments presented next are not a proof of the theorem, but are the basis of our understanding of the relationship between the Gaussian and the discrete superposition model, and motivate the search for a technique to lift a code from the superposition network to the Gaussian network.

Consider the networks in Fig. 1(a) and Fig. 1(c). For simplicity, assume node 1 transmits a symbol $x_1$ and node 2 transmits a symbol $x_2$ (instead of a block of symbols each) from the alphabet for the discrete superposition network. Node 3 receives

$$y'_3 = [[h_{13}]x_1] + [[h_{23}]x_2]$$

in the discrete superposition network in Fig. 1(c), and it receives

$$y_3 = h_{13}x_1 + h_{23}x_2 + z_3$$

in the Gaussian network in Fig. 1(a). Rewriting $y_3$, we get

$$\begin{aligned} y_3 &= y'_3 + (h_{13}x_1 - [h_{13}]x_1) + ([h_{13}]x_1 - [[h_{13}]x_1]) \\ &\quad + (h_{23}x_2 - [h_{23}]x_2) + ([h_{23}]x_2 - [[h_{23}]x_2]) + z_3 \\ &=: y'_3 + v_3 + z_3. \quad (26) \end{aligned}$$

Here we have replaced the actual values of the channel gains with appropriate variables. By definition $y'_3$ lies in $\mathbb{Z} + \imath\mathbb{Z}$. Hence $y'_3$ can be recovered from $y_3$, the quantized values $[v_3]$ and $[z_3]$ respectively of $v_3$ and $z_3$, and the quantized carry $c_3$ obtained from adding the fractional parts of $v_3$ and $z_3$, with

$$c_3 := [(v_3 - [v_3]) + (z_3 - [z_3])]. \quad (27)$$

$[v_3]$ and $[z_3]$ are defined as quantized functions of $v_3$ and $z_3$, as in (1).

So,
$$y'_3 = [y_3] - [v_3] - [z_3] - c_3,$$

and

$$H(y'_3|y_3) \leq H([v_3]) + H([z_3]) + H(c_3), \quad (28)$$

Now, let
$$v_3 = w_{13} + w_{23},$$

where

$$w_{k3} := (h_{k3}x_k - [h_{k3}]x_k) + ([h_{k3}]x_k - [[h_{k3}]x_k]), \ k = 1, 2.$$

Since $|h_{k3} - [h_{k3}]| \leq \sqrt{2}$ and $|x_k| \leq 1$, the magnitude of $v_3$ is less than $2(2\sqrt{2})$. $[v_{3R}]$ and $[v_{3I}]$ lie in $\{-5, -4, \ldots, 5\}$, and $H([v_3]) \leq \log(22)$. The real and imaginary parts of the carry lie in $\{0, \pm 1\}$, hence $H(c_3) \leq 3$. Since $z_3$ is distributed as $\mathcal{CN}(0,1)$, from Lemma 7.1 in the appendix, $H([z_3]) \leq 6$. Adding up all the entropies and substituting in (28), we get the upper bound

$$H(y'_3|y_3) \leq 14.$$

These computations can be repeated for all the nodes in Gaussian network. In general, if there are $M$ incoming signals at a relay node $j$, then the magnitude of $v_j$ is less $M(2\sqrt{2})$, where $v_j$ is similarly defined, as in (26), with respect to the signal received by node $j$. Hence, $v_{jR}$ and $v_{jI}$ will lie in the set $\{-3M + 1, -3M + 2, \ldots, 3M - 1\}$ then

$$H(y'_j|y_j) \leq H([v_j]) + H([z_j]) + H(c_j) \quad (29)$$
$$\leq \log(6M - 1) + 10,$$

where $c_j$ is defined, as in (27), with respect to the signal received by node $j$. Let

$$\kappa := \log(6M - 1) + 10 \quad (30)$$

be a function of the total number of nodes and independent of channel gains (or SNR). Now we use a code designed for the superposition network in the Gaussian network. If there were a genie providing $H(y'_j|y_j)$ bits of data corresponding to the received signal to node $j$ in every channel use, then node $j$ can recover $\underline{\mathbf{y}}'_\mathbf{j}$ from $\underline{\mathbf{y}}_\mathbf{j}$. Since the genie provides at most $\kappa$ bits to every node, it provides a total of at most $M\kappa = \kappa'_\mathcal{R}$ bits per channel use.

Hence, with the genie's aid, a code designed for the discrete superposition network can be used in the Gaussian network at any SNR. Our proof below prunes a fraction of the codewords representing the information that the genie would have provided, so that the decoding can work even without the genie.

### D. Proof of Theorem 3.4

*1) Zero probability of error:* Consider the $(2^{NR}, N)$ code for the superposition network and assume that it has an average probability of error $\delta$, where $0 \leq \delta < 1/2$. Since the superposition network is a noiseless network, each codeword is either always decoded correctly or always decoded incorrectly. Since $\delta < 1/2$, less than half of the codewords are always decoded incorrectly. Discarding them results in a code where all codewords can be successfully decoded, with a small loss in the rate. So, without loss of generality, we assume that the $(2^{NR}, N)$ code (and thus also the $(2^{mNR}, mN)$ code) for the superposition network has zero probability of error.

$\underline{x}_0$, the random variable corresponding to the codeword, has a uniform distribution with $H(\underline{x}_0) = NR$, and induces a distribution on the remaining variables in the network.

*2) Operating over blocks of length $mN$:* In the $(2^{mNR}, mN)$ code, we assume that every node buffers $mN$ of its received symbols, eventually constructing a transmit signal of length $mN$, and transmits it over the next $mN$ channel uses.

For the network in Fig. 1(a), this is possible since nodes can be grouped into levels such that only nodes at one level communicate with another level. For example, nodes 1 and 2 in Fig. 1(a) can buffer their reception till node 0 completes its transmission, then construct their transmit signals, and transmit to nodes 3 and 4 over the next $mN$ channel uses.

For a general network, we need to differentiate between signals received by a node at various time instants to account for causality in construction of their transmit signals. This requires slightly modifying the procedure; see Sec. III-D6.

*3) Pruning the code with respect to node 1:* Each $\underline{\mathbf{y}}'_\mathbf{j}$ (or $\underline{\mathbf{x}}_\mathbf{j}$) in $\mathcal{C}_0$ is generated by $n$ independent samples from the distribution of $\underline{y}'_j$ (or $\underline{x}_j$). Choose $\epsilon > 0$. For a sufficiently large $m$, node 1 has a collection of at most $2^{m(H(\underline{y}'_1)+\epsilon_2)}$ and at least $2^{m(H(\underline{y}'_1)-\epsilon_2)}$ $\epsilon$-strongly typical received vectors in the discrete superposition network corresponding to $\mathcal{C}_0$ (see [35]), where $\epsilon_2 > 0$. As $\epsilon \to 0$, $\epsilon_2 \to 0$. With $\eta$ set to $\epsilon_2$, we construct $\mathcal{S}_1$ by randomly selecting a $2^{-m(N\kappa+2\eta)}$ fraction of this collection. We do this by choosing a subset uniformly among all the subsets of the appropriate size. $|\mathcal{S}_1|$ can be upper bounded as follows (see (29)–(30)):

$$|\mathcal{S}_1| \leq 2^{m(H(\underline{y}'_1)+\epsilon_2)} 2^{-m(N\kappa+2\eta)}$$
$$\leq 2^{m(H(\underline{y}'_1)-H(\underline{y}'_1|\underline{y}_1)-\epsilon_2)} = 2^{m(I(\underline{y}'_1;\underline{y}_1)-\epsilon_2)}.$$

Similarly, we can show that $|\mathcal{S}_1| \geq 2^{m(H(\underline{y}'_1)-N\kappa-3\epsilon_2)}$.

For a large $n$, the number of codewords in $\mathcal{C}_0$ jointly $\epsilon$-strongly typical with a particular $\underline{\mathbf{y}}'_\mathbf{1}$ can be bounded independently of the chosen $\underline{\mathbf{y}}'_\mathbf{1}$; see [35]. The desired set has $2^{m(H(\underline{x}_0|\underline{y}'_1)\pm\epsilon_2)}$ codewords for a particular $\underline{\mathbf{y}}'_\mathbf{1}$, i.e., transmission of one of those codewords in the superposition network results in node 1 receiving the chosen $\underline{\mathbf{y}}'_\mathbf{1}$. Due to the deterministic nature of the channel, the sets of codewords in $\mathcal{C}_0$ jointly typical with two different vectors in $\mathcal{S}_1$ form disjoint sets. To construct $\mathcal{C}_{0,1}$, we pick the set of all codewords in $\mathcal{C}_0$ that are jointly $\epsilon$-strongly typical with some vector in $\mathcal{S}_1$. We have,

$$|\mathcal{C}_{0,1}| = \sum_{\underline{\mathbf{y}}_1' \in \mathcal{S}_1} (\text{\# of codewords in } \mathcal{C}_0 \text{ jointly } \epsilon\text{-strongly typical with } \underline{\mathbf{y}}_1')$$

$$\leq \sum_{\underline{\mathbf{y}}_1' \in \mathcal{S}_1} 2^{m(H(\underline{x}_0|\underline{y}_1') + \epsilon_2)}$$

$$\leq 2^{m(H(\underline{y}_1') - N\kappa - \epsilon_2)} \times 2^{m(H(\underline{x}_0|\underline{y}_1') + \epsilon_2)} \quad (31)$$

$$= 2^{m(H(\underline{x}_0, \underline{y}_1') - N\kappa)} \quad (32)$$

$$= 2^{m(H(\underline{x}_0) - N\kappa)}, \quad (33)$$

where (33) follows since $H(\underline{y}_1'|\underline{x}_0) = 0$. Similarly, we can show that $|\mathcal{C}_{0,1}| \geq 2^{m(H(\underline{x}_0) - N\kappa - 4\epsilon_2)}$.

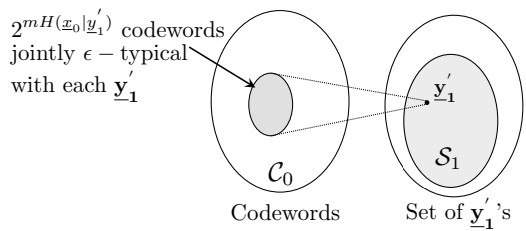

Fig. 4. Pictorial representation of pruning with respect to node 1.

If the source transmits a codeword from $\mathcal{C}_{0,1}$ in the Gaussian network, then the signal $\underline{\mathbf{y}}_1$ received by node 1 can be regarded as a noisy version of the signal $\underline{\mathbf{y}}_1'$ it would have received in the superposition network, as shown in (26). Therefore, we define a channel with input $\underline{y}_1'$ and output $\underline{y}_1$. Node 1 decodes by finding a vector in $\mathcal{S}_1$ that is jointly weakly $\epsilon$-typical with the received vector in the Gaussian network[3]. Since $|\mathcal{S}_1| \leq 2^{m(I(\underline{y}_1';\underline{y}_1) - \epsilon_2)}$, decoding is successful with block error probability less than $\zeta$, where $\zeta \to 0$ as $n \to \infty$.

*4) Further pruning the set of codewords with respect to node 2:* There are $2^{m(H(\underline{y}_2'|\underline{y}_1') \pm \epsilon_2)}$ vectors in the set of $\underline{y}_2'$'s at node 2 that are jointly $\epsilon$-strongly typical with a particular $\underline{y}_1' \in \mathcal{S}_1$. Since we constructed $\mathcal{S}_2$ by randomly choosing a subset containing a $2^{-m(N\kappa + 2\epsilon_2)}$ fraction of the set of all $\underline{y}_2'$'s, for a large $n$, there are $2^{m(H(\underline{y}_2'|\underline{y}_1') - N\kappa \pm 3\epsilon_2)}$ vectors in $\mathcal{S}_2$ jointly $\epsilon$-strongly typical with each $\underline{y}_1' \in \mathcal{S}_1$. Hence, there are $2^{m(H(\underline{y}_1', \underline{y}_2') - 2N\kappa \pm 6\epsilon_2)}$ jointly $\epsilon$-strongly typical vectors in $\mathcal{S}_1 \times \mathcal{S}_2$ with high probability (whp) as $n \to \infty$.

Now, $2^{m(H(\underline{x}_0|\underline{y}_1',\underline{y}_2') \pm \epsilon_2)}$ codewords in $\mathcal{C}_0$ are jointly $\epsilon$-strongly typical with each $\epsilon$-strongly typical tuple in $\mathcal{S}_1 \times \mathcal{S}_2$. We iterate the procedure in the previous subsection by collecting the codewords in $\mathcal{C}_0$ which are jointly $\epsilon$-strongly typical with the $\epsilon$-strongly typical tuples in $\mathcal{S}_1 \times \mathcal{S}_2$, and denote this set by $\mathcal{C}_{0,1,2}$. Naturally, $\mathcal{C}_{0,1,2}$ is a subset of $\mathcal{C}_{0,1}$. As in (31)–(33), we obtain $|\mathcal{C}_{0,1,2}|$ is about $2^{m(H(\underline{x}_0) - 2N\kappa \pm 7\epsilon_2)}$ whp.

If the source transmits a codeword from $\mathcal{C}_{0,1,2}$, then nodes 1 and 2 can correctly decode to vectors in $\mathcal{S}_1$ and $\mathcal{S}_2$ respectively, with high probability for a large $n$, since $|\mathcal{S}_j| \leq 2^{m(I(\underline{y}_j';\underline{y}_j) - \epsilon_2)}$ for $j \in \{1, 2\}$.

---

[3]Since $\underline{\mathbf{y}}_1$ is a continuous signal, we use weak typicality to define the decoding operation. Note that strongly typical sequences are also weakly typical; hence sequences in $\mathcal{S}_1$ are weakly typical.

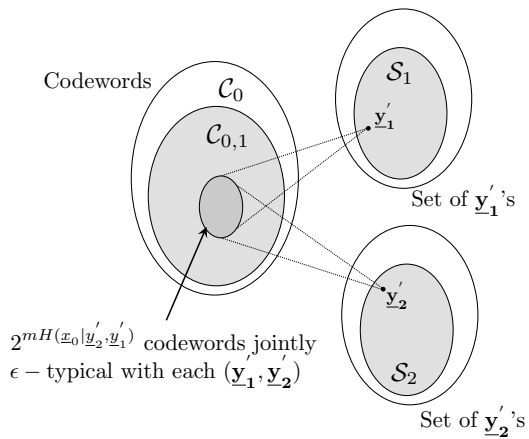

Fig. 5. Pictorial representation of further pruning with respect to node 2.

*5) Further pruning with respect to the remaining nodes:* The same procedure is repeated with respect to the remaining nodes in the network. In the end, we obtain a collection of at most $2^{m(H(\underline{x}_0) - MN\kappa + \epsilon_M)}$ and at least $2^{m(H(\underline{x}_0) - MN\kappa - \epsilon_M)}$ codewords whp, denoted by $\mathcal{C}_{0,1,\cdots,M} =: \mathcal{C}_G$, where $\epsilon_M > 0$. Note that $\epsilon_M \to 0$ as $\epsilon \to 0$. Transmission of a codeword in $\mathcal{C}_G$ results in the received signal at node $j$ in the superposition network belonging to the set $\mathcal{S}_j$.

Now, if $\mathcal{C}_G$ is used on the Gaussian network with encoding and decoding procedures at all nodes as described above, then the destination can decode to the transmitted codeword whp. Thus, on the Gaussian network, $\mathcal{C}_G$ achieves the rate

$$H(\underline{x}_0)/N - M\kappa - \epsilon_M/M = R - M\kappa - \epsilon_M/M$$
$$= R - \kappa_\mathcal{R}' - \epsilon_M/M, \quad (34)$$

where $\epsilon_M$ can be made arbitrarily small.

*6) Interleaving the codewords for general networks:* As mentioned in Sec. III-D2, we need to slightly modify the lifting procedure for relay networks which have irregular level sets that do not permit straightforward buffering of received symbols at a node.

In this case, codewords in $\mathcal{C}_0$ are constructed by adjoining $N$ blocks of $m$ symbols each, where the first block $\underline{x}_0(1)$ consists only of the first symbols of $m$ codewords of the $(2^{NR}, N)$ code, the second block $\underline{x}_0(2)$ consists only of the second symbols of the same codewords, and so on. The source transmits $\underline{x}_0(t)$'s in the order of increasing $t$.

In the $(2^{NR}, N)$ code, let $y_j'(t)$, $t = 1, \ldots, N$, denote the $t$-th symbol received by node $j$. We adjoin the $t$-th received symbols from $m$ uses of the code to construct $\underline{y}_j'(t)$. Since $x_j(t)$, the $t$-th symbol transmitted by node $j$, is a function of $\{y_j'(p)\}_{p=1}^{t-1}$, node $j$ can construct $\underline{x}_j(t)$, vector consisting of the $t$-th transmit symbols from $m$ uses of the code, after receiving $\{\underline{y}_j'(p)\}_{p=1}^{t-1}$, .

Essentially, we interleave the symbols from $m$ uses of the same code to ensure that the nodes can buffer their receptions.

In order to lift the coding scheme to the Gaussian network, we prune $\mathcal{C}_0$ by randomly picking a $2^{-m(\kappa + 2\eta)}$-fraction of the set of $\epsilon$-strongly typical $\underline{y}'_j(t)$, for all $t$, for all $j$, and collecting the codewords jointly $\epsilon$-strongly typical with them to form $\mathcal{C}_G$.

In the Gaussian network, each node buffers its reception for $m$ time units, decodes to the appropriate $\underline{y}'_j(t)$, constructs $\underline{x}_j(t+1)$, transmits it on the next $m$ time units. The destination decodes individual $m$-length blocks to get $\underline{y}'_M(t)$, $t = 1, 2, \ldots, N$, and decodes to a codeword in $\mathcal{C}_G$ after de-interleaving $\{\underline{y}'_M(t)\}$.

This completes the proof of Theorem 3.4.

## IV. A DIGITAL INTERFACE FOR GAUSSIAN INTERFERENCE CHANNELS

In this section we show that the capacity regions of the Gaussian interference channel and the discrete superposition interference channel are within a bounded number of bits, independent of channel gains or SNR. This result was proved for the case of $2 \times 2$ interference channel (and a slightly different deterministic model) in [8], and we use some of the techniques here. Also, similar to the case of the relay network, we develop a systematic way to 'lift' any code for the discrete superposition interference network to the Gaussian interference network, and establish that it does so with no more than a bounded loss in the rate.

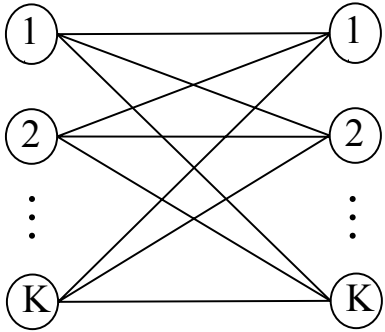

Fig. 6. $K \times K$ interference network.

Consider the $K \times K$ Gaussian interference network shown in Fig. 6. There are $K$ sources, labeled $1, 2, \ldots, K$, who want to transmit data to $K$ destinations, labeled $1, 2, \ldots, K$. All the sources are independent of each other, and the $k$-th source wants to transmit information only to the $k$-th destination. At the $k$-th destination, the transmissions by other sources interfere with the transmission of the $k$-th source. Let $x_i$ be the signal transmitted by the $i$-th source, for $i = 1, 2, \ldots, K$. The received signal at the $j$-th destination is

$$y_j = \sum_{k=1}^{K} h_{kj} x_k + z_j, \quad j = 1, 2, \ldots, K, \quad (35)$$

where $z_j$ is $\mathcal{CN}(0, 1)$.

Consider the corresponding discrete superposition model for this interference channel. The received signal at the $j$-th destination in the discrete superposition model is

$$y_j = \sum_{k=1}^{K} [[h_{kj}] x_k], \quad j = 1, 2, \ldots, K, \quad (36)$$

where $[h_{kj}]$ is the quantized version of $h_{kj}$.

Consider a block code for the interference network, either the Gaussian version or the discrete superposition version. Such a $(2^{NR_1}, 2^{NR_2}, \ldots, 2^{NR_K}, N)$ code for the interference channel is defined by an encoding function for each source

$$\underline{x}_k : \{1, 2, \ldots, 2^{NR_k}\} \to \mathcal{X}^N, \text{ for } k = 1, 2, \ldots, K,$$

where $\mathcal{X}$ is the input alphabet of the channel, and a decoding function for each destination

$$g_k : \mathcal{Y}_k^N \to \{1, 2, \ldots, 2^{NR_k}\}, \text{ for } k = 1, 2, \ldots, K,$$

where $\mathcal{Y}_k$ is the output alphabet of the channel at destination node $k$. Let $\mathcal{M}_j$ be a random variable uniformly distributed on $\{1, 2, \ldots, 2^{NR_j}\}$, for each $j$, corresponding to the message that source $j$ wants to communicate to destination $j$. $\mathcal{M}_j$ is mapped to the codeword $\underline{x}_j(\mathcal{M}_j)$. The average probability of error is given by

$$P_e = Pr(g_k(\underline{y}_k) \neq \mathcal{M}_k, \text{ for some } k),$$

where $\underline{y}_k$ is the signal received by destination $k$. The capacity region is the collection of all rate tuples $\underline{R}$ such that for any $\epsilon > 0$, there exists a blocklength $N$ for which $P_e < \epsilon$.

All rate vectors in the capacity region are referred to as achievable rate vectors. The next lemma states the equivalence between achievable rate vectors and a collection of multi-letter mutual information terms.

*Lemma 4.1:* A rate vector $\underline{R}$ lies in the capacity region of the $K \times K$ interference channel if and only if there exists a blocklength $N$, and a collection of independent random variables $\{\underline{x}_k, k = 1, 2, \ldots, K\}$ such that

$$R_k < \frac{1}{N} I(\underline{x}_k; \underline{y}_k), \text{ for } k = 1, 2, \ldots, K, \quad (37)$$

where $\underline{y}_k$ is the received signal at the $k$-th receiver.

*Proof:* The proof of this lemma is similar to the proof of Lemma 3.1, and we need to consider the rate achieved by every source-destination pair individually. We skip the details. ∎

The main result of this section is:

*Theorem 4.2:* Consider the $K \times K$ interference channel described above. The capacity region of the Gaussian interference channel and the discrete superposition interference channel are within a bounded gap, where the gap is independent of channel gains or SNR.

If $\underline{R}_G$ is a rate vector in the capacity region of the Gaussian interference channel, then there is a rate vector $\underline{R}_D$ in the capacity region of the discrete superposition interference channel such that

$$|R_{k,G} - R_{k,D}| \leq \kappa_\mathcal{I}, \text{ for } k = 1, 2, \ldots, K, \quad (38)$$

where

$$\kappa_\mathcal{I} := 6K + \log(144K + 1). \quad (39)$$

Conversely, a coding scheme for the discrete superposition interference channel corresponding to a rate vector $\underline{R}_D$ can be lifted to the Gaussian interference channel to obtain an achievable rate vector $\underline{R}_G$ with a loss of at most $\kappa'_\mathcal{I}$ bits, where

$$\kappa'_\mathcal{I} := \log(6K-1) + 10. \quad (40)$$

We prove the theorem in a series of steps. First, using Lemma 4.1, we convert any achievable rate tuple for either the Gaussian or discrete superposition network into a set of mutual information terms. Then, by analyzing these mutual information terms, we prove that any coding scheme for the Gaussian channel can be transformed into a coding scheme for the discrete superposition channel, with at most a bounded loss in the rate. Similarly, we prove that any coding scheme for the discrete superposition interference channel (DSIC) can be lifted to the Gaussian interference channel (GIC), again with at most a bounded loss in the rate.

### A. Capacity(DSIC) ⊆ Capacity(GIC), to within an additive constant

To begin, we prove that the capacity region of the discrete superposition interference network is at least as large as that of the Gaussian interference network, minus a constant number of bits.

*Lemma 4.3:* Let $\underline{R}_G$ be a vector in the capacity region of the Gaussian interference channel. Then, there exists a rate vector $\underline{R}_D$ in the capacity of the discrete superposition interference channel such that

$$R_{Dk} \geq R_{Gk} - \kappa_\mathcal{I}, \text{ for } k=1,2,\ldots,K, \quad (41)$$

with $\kappa_\mathcal{I}$ defined in (39).

*Proof:* From Lemma 4.1, we know that we can replace any achievable rate vector $\underline{R}_G$ by a collection of mutual information terms $\{I(\underline{x}_k; \underline{y}_k), k=1,2,\ldots,K\}$. We perform a series of approximations to convert the mutual information $I(\underline{x}_j; \underline{y}_j)$ corresponding to the rate of the $j$-th source to a mutual information term for the discrete superposition interference channel, incurring a loss no more than a bounded amount. Rest of the arguments are similar to the proof in Section III-A.

The received signal at the $j$-th receiver in the Gaussian interference channel is

$$\underline{y}_j = \sum_{k=1}^{K} h_{kj} \underline{x}_k + \underline{z}_j, \quad (42)$$

with the transmitted signals satisfying an average unit power constraint.

Note that $\underline{x}_k = (x_{k1}, x_{k2}, \ldots, x_{kN})$, and each $x_{km}$ can be split into its quantized part $[x_{km}]$ and fractional part $\tilde{x}_{km}$, where

$$\tilde{x}_{km} := x_{km} - [x_{km}].$$

We discard $[x_{km}]$ and retain $\tilde{x}_{km}$. Since $x_{km}$ satisfies a unit average power constraint, $\tilde{x}_{km}$ also satisfies a unit average power constraint. Define

$$\underline{\tilde{y}}_j := \sum_{k=1}^{K} h_{kj} \underline{\tilde{x}}_k + \underline{z}_j. \quad (43)$$

Denote the discarded portion of the received signal by

$$\underline{\hat{y}}_j := \sum_{k=1}^{K} h_{kj} [\underline{x}_k]. \quad (44)$$

Comparing the mutual information corresponding to the $j$-th source-destination pair for channels (42) and (43), we get

$$\begin{aligned}
NR_{Gj} &= I(\underline{x}_j; \underline{y}_j) \\
&\leq I(\underline{x}_j; \underline{\tilde{y}}_j, \underline{\hat{y}}_j) \\
&\leq I(\underline{\tilde{x}}_j, [\underline{x}_j]; \underline{\tilde{y}}_j, \underline{\hat{y}}_j) \\
&= I(\underline{\tilde{x}}_j, [\underline{x}_j]; \underline{\tilde{y}}_j) + I(\underline{\tilde{x}}_j, [\underline{x}_j]; \underline{\hat{y}}_j | \underline{\tilde{y}}_j) \\
&= I(\underline{\tilde{x}}_j; \underline{\tilde{y}}_j) + I(\underline{\tilde{x}}_j, [\underline{x}_j]; \underline{\hat{y}}_j | \underline{\tilde{y}}_j) \quad (45) \\
&\leq I(\underline{\tilde{x}}_j; \underline{\tilde{y}}_j) + H(\underline{\hat{y}}_j) \\
&\leq I(\underline{\tilde{x}}_j; \underline{\tilde{y}}_j) + \sum_{k=1}^{K} H([\underline{x}_k]), \quad (46)
\end{aligned}$$

where (45) follows because $[\underline{x}_j] \to \underline{\tilde{x}}_j \to \underline{\tilde{y}}_j$ form a Markov chain, and (46) holds because $\hat{y}_{jn}$ is a function of $\{[x_{kn}]\}$ from (44). It is proved in the appendix that $H([x_{km}]) \leq 6$, hence we get

$$I(\underline{\tilde{x}}_j; \underline{\tilde{y}}_j) \geq I(\underline{x}_j; \underline{y}_j) - 6KN. \quad (47)$$

Since $\tilde{x}_{kmR}$ and $\tilde{x}_{kmI}$ lie in $(-1,1)$, we obtain positive inputs by adding 1 to each. This is equivalent to adding $\sum_k h_{kj}(1+\imath)$ to $\tilde{y}_{jm}$. Denoting by $\underline{\nu}$ the vector of $(1+\imath)$'s, we also divide by $2\sqrt{2}$ throughout to get:

$$\begin{aligned}
(\underline{\tilde{y}}_j + \sum_k h_{kj}\underline{\nu})/2\sqrt{2} \\
= \sum_k h_{kj}(\underline{\tilde{x}}_k + \underline{\nu})/2\sqrt{2} + \underline{z}_j/2\sqrt{2}. \quad (48)
\end{aligned}$$

Note that

$$\begin{aligned}
I(\underline{\tilde{x}}_j; \underline{\tilde{y}}_j) &= I\left(\underline{\tilde{x}}_j, \frac{\underline{\tilde{x}}_j + \underline{\nu}}{2\sqrt{2}}; \underline{\tilde{y}}_j\right) \\
&= I\left(\underline{\tilde{x}}_j, \frac{\underline{\tilde{x}}_j + \underline{\nu}}{2\sqrt{2}}; \underline{\tilde{y}}_j, \frac{\underline{\tilde{y}}_j + \sum_k h_{kj}\underline{\nu}}{2\sqrt{2}}\right) \\
&= I\left(\frac{\underline{\tilde{x}}_j + \underline{\nu}}{2\sqrt{2}}; \frac{\underline{\tilde{y}}_j + \sum_k h_{kj}\underline{\nu}}{2\sqrt{2}}\right).
\end{aligned}$$

To avoid introducing new notation, for the rest of the proof we abuse notation and denote the left hand side of (48) by $\underline{y}_j$, $(\underline{\tilde{x}}_k + \underline{\nu})/2\sqrt{2}$ by $\underline{x}_k$, and $\underline{z}_j/2\sqrt{2}$ by $\underline{z}_j$, for all $j$.

With the new notation, $|x_{km}| \leq 1$, with positive real and imaginary parts, and $z_{jm}$ is distributed as $\mathcal{CN}(0, 1/8)$.

The features of the model that we next address are:
1) channel gains are quantized to lie in $\mathbb{Z} + \imath\mathbb{Z}$,
2) real and imaginary parts of the scaled inputs are restricted to

$$n := \max_{(i,j)\in\mathcal{E}} \max\{\lfloor\log|h_{ijR}|\rfloor, \lfloor\log|h_{ijI}|\rfloor\}$$

bits,
3) there is no AWGN, and
4) outputs are quantized to lie in $\mathbb{Z} + \imath\mathbb{Z}$.

Let the binary expansion of $\sqrt{2}\,x_{kmR}$ be $0.x_{kmR}(1)x_{kmR}(2)\ldots$, i.e.,

$$x_{kmR} =: \frac{1}{\sqrt{2}}\sum_{p=1}^{\infty} 2^{-p} x_{kmR}(p).$$

The received signal in the discrete superposition channel only retains the following relevant portion of the input signals:

$$\underline{y}'_j = \sum_k [[h_{kj}]\,\underline{x}'_k], \qquad (49)$$

where

$$x'_{kmR} := \frac{1}{\sqrt{2}}\sum_{p=1}^{n} x_{kmR}(p) 2^{-p},$$

$$x'_{kmI} := \frac{1}{\sqrt{2}}\sum_{p=1}^{n} x_{kmI}(p) 2^{-p},$$

with $n$ defined in (3). To obtain (49) we subtracted $\underline{\delta}_j$ from $\underline{y}_j$, where

$$\begin{aligned}\underline{\delta}_j &:= \sum_{k=1}^{K} \Big( h_{kj}(\underline{x}_k - \underline{x}'_k) + (h_{kj} - [h_{kj}])\,\underline{x}'_k + \quad (50) \\ &\qquad ([h_{kj}]\,\underline{x}'_k - [[h_{kj}]\,\underline{x}'_k])\Big) + \underline{z}_j \\ &=: \sum_k \underline{w}_{kj} + \underline{z}_j \\ &=: \underline{v}_j + \underline{z}_j,\end{aligned}$$

To bound the loss in the mutual information in the discrete superposition network from the original Gaussian interference network, we have

$$\begin{aligned}I(\underline{x}_j;\underline{y}_j) &\le I(\underline{x}_j;\underline{y}_j,\underline{y}'_j,\underline{\delta}_j) \\ &= I(\underline{x}_j;\underline{y}'_j,\underline{\delta}_j) \\ &= I(\underline{x}_j;\underline{y}'_j) + I(\underline{x}_j;\underline{\delta}_j|\underline{y}'_j) \\ &= I(\underline{x}'_j;\underline{y}'_j) + h(\underline{\delta}_j|\underline{y}'_j) - h(\underline{\delta}_j|\underline{y}'_j,\underline{x}_j) \\ &\le I(\underline{x}'_j;\underline{y}'_j) + h(\underline{\delta}_j) - h(\underline{\delta}_j|\underline{y}'_j,\underline{x}_j,\underline{v}_j) \\ &= I(\underline{x}'_j;\underline{y}'_j) + h(\underline{\delta}_j) - h(\underline{z}_j) \\ &= I(\underline{x}'_j;\underline{y}'_j) + I(\underline{v}_j;\underline{\delta}_j).\end{aligned}$$

By bounding the magnitudes of the terms in (50), we get $|w_{kjm}| \le 3\sqrt{2}$. So, $I(\underline{v}_j;\underline{\delta}_j)$ is the mutual information of $N$ uses of a Gaussian MISO channel with average input power constraint less than $(3\sqrt{2})^2 K \le 18K$ and

$$\begin{aligned}I(\underline{v}_j;\underline{\delta}_j) &\le N \log(1 + 18K/(1/8)) \\ &< N \log(1 + 144K).\end{aligned} \qquad (51)$$

Note that $I(\underline{x}'_j;\underline{y}'_j)$ is the mutual information between the input and output of the $j$-th source-destination pair in the discrete superposition interference channel. By Lemma 4.1, this mutual information translates into an achievable rate $R_{Dj} = I(\underline{x}'_j;\underline{y}'_j)/N$. By accumulating the losses in transforming the inputs for the Gaussian channel into the corresponding inputs for the deterministic channel, we obtain

$$R_{Dj} \ge R_{Gj} - (6K + \log(1+144K)), \qquad (52)$$

thereby proving the lemma. ∎

*B. Capacity(GIC) ⊆ Capacity(DSIC), to within an additive constant*

Next we prove that the capacity region of the Gaussian interference channel is at least as large as that of the discrete superposition interference channel, to within an additive constant. Once again, we prove this by converting a coding scheme for the discrete superposition interference channel to a mutual information expression using Lemma 4.1, and bounding the loss in the mutual information when transforming the discrete superposition model to the Gaussian model. The proof of the lemma below contains all the ideas on lifting a code from the discrete superposition model to the Gaussian model, and we explicitly mention them in a subsequent subsection.

*Lemma 4.4:* Let $\underline{R}_D$ be a vector in the capacity region of the discrete superposition interference channel. Then, there exists a rate vector $\underline{R}_G$ in the capacity of the Gaussian interference channel such that

$$R_{Gk} \ge R_{Dk} - \kappa'_{\mathcal{I}}, \qquad (53)$$

for $k = 1, 2, \ldots, K$, where $\kappa'_{\mathcal{I}}$ is defined in (40).

*Proof:* Pick any coding scheme for the discrete superposition interference channel which achieves the rate tuple $R_D$. From Lemma 4.1, we know that this corresponds to a collection of mutual information terms $\{I(\underline{x}_k;\underline{y}'_k),\ k = 1,2,\ldots,K\}$. Note that here $\underline{x}_k$ corresponds to the input of the $k$-th source in the discrete superposition interference network, and $\underline{y}_k$ corresponds to the output of the $k$-th destination in the discrete superposition network. Now, we use the same input distribution on the Gaussian interference network and show that the mutual information corresponding to the rate of the $j$-th source does not decrease by more than a bounded amount. The same arguments will be applicable to the remaining mutual information terms.

Since every element of the vector $\underline{x}_k$ satisfies a peak power constraint, it also satisfies an average power constraint. Hence it can be used as an input to the Gaussian channel to get

$$\underline{y}_j = \sum_k h_{kj}\underline{x}_k + \underline{z}_j. \qquad (54)$$

Since we know that

$$\underline{y}'_j = \sum_k [[h_{kj}]\underline{x}_k], \qquad (55)$$

we have

$$\begin{aligned}\underline{y}_j &=: \underline{y}'_i + \sum_{k=1}^{K} \underline{w}_{kj} + \underline{z}_j \\ &=: \underline{y}'_i + \underline{v}_j + \underline{z}_j,\end{aligned} \qquad (56)$$

where $\underline{w}_{kj}$ is defined as

$$\underline{w}_{kj} := (h_{kj} - [h_{kj}])\,\underline{x}_k + ([h_{kj}]\,\underline{x}_k - [[h_{kj}]\,\underline{x}_k]).$$

By definition $\underline{y}'_j$ is a vector of entries from $\mathbb{Z} + \imath\mathbb{Z}$. Hence $\underline{y}'_j$ can be recovered from $\underline{y}_j$, the quantized values of $\underline{v}_j$ and $\underline{z}_j$, and the vector of carries $\underline{c}_j$ obtained from adding the fractional parts of $\underline{v}_j$ and $\underline{z}_j$. So,

$$\begin{aligned} I(\underline{x}_j; \underline{y}'_j) & \\ \leq\ & I(\underline{x}_j; \underline{y}_j, [\underline{v}_j], [\underline{z}_j], \underline{c}_i) \\ \leq\ & I(\underline{x}_j; \underline{y}_j) + H([\underline{v}_j]) + H([\underline{z}_j]) + H(\underline{c}_j), \quad (57) \end{aligned}$$

where $[\underline{v}_j]$ and $[\underline{z}_j]$ are defined as quantized functions of $\underline{v}_j$ and $\underline{z}_j$. Similar to the arguments preceding (29), we get

$$\begin{aligned} H([\underline{v}_j]) &\leq N \log(12K - 2), \\ H(\underline{c}_j) &\leq 3N, \\ H([\underline{z}_j]) &\leq 6N. \end{aligned}$$

In (57), $I(\underline{x}_j; \underline{y}_j)$ corresponds to $N$ times the rate $R_{Gj}$ achieved on the Gaussian interference channel by the $j$-th source. Therefore we get

$$R_{Gj} \geq R_{Dj} - \log(6K - 1) - 10.$$

∎

Lemmas 4.3 and 4.4 together complete the proof of Theorem 4.2.

*C. Lifting codewords to the Gaussian interference network*

Since the construction of a digital interface and the procedure of lifting codewords from the discrete superposition interference channel to the Gaussian interference channel was implicit in the proof of the above lemma, we summarize the procedure below:

- Consider any coding scheme for the discrete superposition interference channel that achieves a rate tuple $\underline{R}_D$ with probability of error $\epsilon$. Using Lemma 4.1, it can be converted to a collection of mutual information terms $\{I(\underline{x}_k; \underline{y}'_k)\}$, where the $k$-th term corresponds to the rate of the $k$-th source-destination pair, and a collection of independent input distributions $\{p(\underline{x}_k)\}$, where the $k$-th distribution corresponds to the input distribution for the $k$-th source. The input distribution $p(\underline{x}_k)$ is over $N$-length vectors, where $N$ is the length of the codewords for the discrete superposition interference channel.
- We construct a $mN$-length codeword for the $k$-th source by picking $m$ vectors independently with the distribution $p(\underline{x}_k)$, and adjoining them. We construct $2^{mN(R_{Dj}-\kappa'_\mathcal{I})}$ codewords this way. This set forms the $(2^{mN(R_{Dj}-\kappa'_\mathcal{I})}, mN)$ code for the $j$-th source in the Gaussian interference network. Similar to the lifting procedure for the relay network, we can visualize the construction of codewords in the $(2^{mN(R_{Dj}-\kappa'_\mathcal{I})}, mN)$ code by referring to Fig. 3.
- In the Gaussian interference channel, with joint typical decoding, the $k$-th decoder can recover the codeword transmitted by the $k$-th source with probability of error less than $\epsilon$, as we allow $m$ to tend to $\infty$. This is essentially proved in Lemma 4.4.

Therefore, we can operate the Gaussian interference channel on the digital interface defined by the signals transmitted and received in the discrete superposition interference channel. We can use techniques similar to the proof of Theorem 3.4 to prove Lemma 4.4, but we chose to present the simpler proof involving manipulating mutual information terms directly. From Lemmas 4.3 and 4.4, we know that the capacities of both the networks are within a bounded gap. Hence, if we choose a near capacity-achieving coding scheme for the discrete superposition interference network, we can transform it by following the above procedure for lifting codewords and obtain a near-optimal digital interface for operating the Gaussian interference network.

## V. DIGITAL INTERFACES FOR SOME OTHER NETWORKS

In this section, we list some other networks for which the discrete superposition model provides a near-optimal digital interface, and where codes can be lifted from the discrete superposition counterpart to the Gaussian network.

*A. MIMO networks*

From the results in [1] and [2], it is easy to see that MIMO channels are well-approximated by the discrete superposition model in the capacity sense, and we can lift codes from a MIMO discrete superposition channel to a Gaussian MIMO channel. This correspondence extends to more general MIMO networks too.

*1) MIMO relay networks:* MIMO relay networks can be handled in the same way as in Section III, by treating each transmitted/received signal as a collection of vectors, where the size of the collection depends on the number of transmit/receive antennas.

For simplicity, consider a relay network where every node has two transmit and two receive antennas. All transmitted and received signals are a pair of vectors in both the Gaussian and discrete superposition model for this network. For example, node $j$'s received signal in the Gaussian model is $\underline{\mathbf{y}}_\mathbf{j} = [\underline{\mathbf{y}}_{\mathbf{j},\mathbf{1}}, \underline{\mathbf{y}}_{\mathbf{j},\mathbf{2}}]$. The channel on a particular wireless link $(i,j)$ is specified by four channel gains, $\{h_{ij}^{k,l}\}$, where $k \in \{1,2\}$ indexes the transmit antennas of $i$ and $l \in \{1,2\}$ indexes the receive antennas of $j$. Assuming that the noises at both receive antennas are distributed as $\mathcal{CN}(0,1)$, we have

$$\underline{\mathbf{y}}_{\mathbf{j},\mathbf{l}} = \sum_{i \in \mathcal{N}(j)} (h_{ij}^{1,l} \underline{\mathbf{x}}_{\mathbf{i},\mathbf{1}} + h_{ij}^{2,l} \underline{\mathbf{x}}_{\mathbf{i},\mathbf{2}}) + z_{j,l}, \ l = 1,2. \quad (58)$$

We state the counterpart of Theorems 3.2 and 3.4.

*Theorem 5.1:* Consider a relay network where every node has a maximum of $L$ transmit or receive antennas. The capacity $C_G$ of the Gaussian relay network and the capacity $C_D$ of the discrete superposition relay network is within a bounded gap of $\kappa_{\mathcal{R},L}$ bits where

$$\kappa_{\mathcal{R},L} = O(LM \log(LM)). \quad (59)$$

Furthermore, a coding scheme for the discrete superposition MIMO relay network can be lifted to the Gaussian MIMO relay network with a loss of $\kappa'_{\mathcal{R},L}$ bits in the rate, where

$$\kappa'_{\mathcal{R},L} := LM(\log(6LM - 1) + 10). \quad (60)$$

*Proof:* The arguments in Section III unchanged to this model by replacing every vector with its corresponding tuple of vectors. A simple way to derive the above results is to replace each set of $L$ antennas at a transmitter or receiver by a collection of $L$ virtual nodes. Now there are a total of $LM$ virtual nodes in the network. The bound in (59) follows by replacing $M$ with $LM$ in Theorem 3.2. In order to lift the code from the discrete superposition MIMO network to the Gaussian relay network, we need to prune the source's codebook with respect to all the virtual nodes. Hence (60) follows by replacing $M$ in (25) with $LM$. ∎

*2) MIMO interference networks:* The results in Theorem 4.2 can be extended to $K \times K$ interference channels where each transmitter and destination have multiple antennas. Once again, the constant determining the bounded gap is a function of the number of nodes in the network, as well as the number of transmit and receive antennas at the various nodes.

*Theorem 5.2:* Consider the $K \times K$ MIMO interference channel described above, where every node has a maximum of $L$ antennas. The capacity region of the Gaussian MIMO interference channel and the discrete superposition MIMO interference channel are within a bounded gap of $\kappa_{\mathcal{I},L}$ bits, where

$$\kappa_{\mathcal{I},L} := 6LK + L\log(144LK + 1). \quad (61)$$

*Furthermore, a coding scheme for the discrete superposition MIMO interference channel can be lifted to the Gaussian MIMO interference channel with a loss of $\kappa'_{\mathcal{I},L}$ bits in the rate, where*

$$\kappa'_{\mathcal{I},L} := L(\log(6LK - 1) + 10). \quad (62)$$

*Proof:* The proof of this theorem is similar to that of Theorem 4.2. We can treat each transmitting antenna and receiving antenna as a virtual node. For proving (61), we compare with (46) and note that in this case there are at most $KL$ virtual transmitters that contribute $6LK$ to the bound. Since there are $L$ virtual receivers at every node and each one contributes $\log(144LK+1)$ to the bound, where the arguments are similar to those preceding (51), the total contribution of the virtual receivers is $L\log(144LK + 1)$. Adding up the two contributions, we get the bound in (61).

The bound in (62) can be proved using the techniques in Lemma 4.4. Each of the $L$ virtual receivers at a particular receiver contributes $\log(6LK - 1) + 10$ to the bound, with a total of contribution of $L(\log(6LK - 1) + 10)$ due to all the virtual receivers at a node. ∎

### B. Multicast

Consider a relay network with $M+1$ nodes, with the nodes labeled as $0, 1, \ldots, M$, where node $0$ is the source and it wants to communicate the same information to a subset of the remaining nodes. The other nodes which are not the intended recipients act as relays. It is known that the cut-set bound is the capacity of such multicast networks, up to an additive constant [3], [7].

*Theorem 5.3:* Consider a multicast relay network mentioned above. The capacity $C_G$ of the Gaussian relay network and the capacity $C_D$ of the discrete superposition relay network is within a bounded gap of $\kappa_{\mathcal{M}}$ bits where

$$\kappa_{\mathcal{M}} = O(M\log(M)). \quad (63)$$

*Furthermore, a coding scheme for the discrete superposition multicast relay network can be lifted to the Gaussian multicast relay network with a loss of $\kappa'_{\mathcal{M}}$ bits in the rate, where*

$$\kappa'_{\mathcal{M}} := M(\log(6M - 1) + 10). \quad (64)$$

*Proof:* From the results in [3] and [1], it is easy to prove the bound in (63). The basic idea is that the cut-set bound is approximately achievable for Gaussian and discrete superposition networks. Hence we need to prove that the cut-set bounds for the Gaussian network and the discrete superposition network are within a bounded gap. This is proved in Theorem 3.2.

For proving the results in (64), we choose a coding scheme for the multicast discrete superposition network. The key to lifting the coding scheme to the Gaussian network is that all the intended destinations in a multicast network are decoding the same data. Hence if we prune the source's codebook with respect to all the nodes, as in the proof of Theorem 3.4, then the pruned code can be decoded on the Gaussian network. We skip the details.

We can extend the above theorem to the case when the nodes have multiple transmit and receive antennas. ∎

## VI. CONCLUDING REMARKS

One of the main problems in network information theory is computing the capacity region of a large network with many sources, many destinations, and arbitrary data transmission requirements. Currently, this looks like no more than a distant possibility. As suggested by [7], a possibly simpler aim is to approximate a general network with a deterministic model, perhaps with the discrete superposition model. We have proved that the discrete superposition model serves as a near-optimal digital interface for designing codes for the Gaussian relay network and the Gaussian interference network. This transforms the problem of designing near-optimal codes for the Gaussian network to designing near-optimal codes for the discrete counterpart. Also, the problem of computing the capacity of the Gaussian network is reduced to a combinatorial problem of computing the capacity of a discrete network. In case of the relay network, even though we already know near-optimal coding schemes for the network, it may still be helpful to construct simple codes for its discrete superposition counterpart. Such simple schemes can be directly translated, via the lifting procedure proposed earlier, to construct simple coding schemes for the original Gaussian network. In the case of the Gaussian interference network, computing the capacity region of the $3 \times 3$ discrete superposition interference channel will yield the capacity region of the original Gaussian network to within a constant, and will improve our understanding of larger practical interference networks. We have not been able to prove the near-optimality of the discrete superposition model in approximating the capacity of a general Gaussian network.

A better understanding of the limits of approximating Gaussian networks with noiseless deterministic models will help us in computing the fundamental limits of wireless networks with many users and may also help in designing coding schemes for them.

## VII. APPENDIX

### A. Maximum entropy under a unit power constraint

*Lemma 7.1:* Let $x$ be a random variable whose domain is $\mathbb{Z} + \imath\mathbb{Z}$, with $E[|x|^2] \leq 1$. Then the entropy of $x$ is bounded as $H(x) \leq 6$.

*Proof:* The entropy of $x$ can be bounded as
$$\begin{aligned} H(x) &= H(x_R + \imath x_I) \\ &\leq H(x_R) + H(x_I). \end{aligned}$$

Now
$$\begin{aligned} H(x_R) &= H(\text{sign}(x_R), |x_R|^2) \\ &\leq H(\text{sign}(x_R)) + H(|x_R|^2) \\ &\leq 1 + H(|x_R|^2). \end{aligned}$$

Let $z = |x_R|^2$. The domain of $z$ is the set of non-negative integers. Since $E[|x|^2] \leq 1$, and $E[z] = E[|x_R|^2] \leq 1$. If $E[z] < 1$, then $p(z = 0) > 0$. Since the alphabet of $z$ is countably infinite, there is always a $k \in \mathbb{Z}^+$ with $p(z = k) < p(z = 0)$, with the LHS possibly zero. Now mixing the probability distribution of $z$ by replacing each of $p(z = 0)$ and $p(z = k)$ by the average of the two probabilities will increase the entropy (see [35]), and will also increase the mean. Hence the entropy of $z$, subject to $E[z] \leq 1$, is maximized when $E[z] = 1$. For a given mean, the geometric distribution maximizes the entropy among all discrete distributions (see [35]). Since the entropy of a geometric random variable over the non-negative integers with unit mean is 2, $H(|x_R|^2) = H(z) \leq 2$.

Hence $H(x_R) \leq 3$. Similarly, we can prove that $H(x_I) \leq 3$. Combining the two bounds, we get the statement of the lemma. ∎

**M. Anand** Anand obtained his B.E. from VTU, Belgaum, in 2003, and M.S. from IISc, Bangalore, in 2007, all in ECE. He is a PhD student in ECE, UIUC, since Fall 2007 and his advisor is Prof. P. R. Kumar.

Previously, he was an engineer in the voiceband modem team at Ittiam Systems, Bangalore, from June 2003 - May 2004. He was an intern in the systems team at Qualcomm Flarion Tech in summer 2010 where he worked on P2P over licensed spectrum.

His interests are broadly in wireless systems, networks, and stochastic control.

**P. R. Kumar** P. R. Kumar obtained his B. Tech. degree in Electrical Engineering (Electronics) from I.I.T. Madras in 1973, and the M.S. and D.Sc. degrees in Systems Science and Mathematics from Washington University, St. Louis, in 1975 and 1977, respectively. From 1977-84 he was a faculty member in the Department of Mathematics at the University of Maryland Baltimore County. Since 1985 he has been at the University of Illinois, Urbana-Champaign, where he is currently Franklin W. Woeltge Professor of Electrical and Computer Engineering, Research Professor in the Coordinated Science Laboratory, Research professor in the Information Trust Institute, and Affiliate Professor of the Department of Computer Science. He has worked on problems in game theory, adaptive control, stochastic systems, simulated annealing, neural networks, machine learning, queueing networks, manufacturing systems, scheduling, wafer fabrication plants and information theory. His current research interests are in wireless networks, sensor networks, and networked embedded control systems. He has received the Donald P. Eckman Award of the American Automatic Control Council, the IEEE Field Award in Control Systems, and the Fred W. Ellersick Prize of the IEEE Communications Society. He is a Fellow of the IEEE, a member of the US National Academy of Engineering, and an Associate Fellow of the Academy of Sciences for the Developing World. He has been awarded an honorary doctorate by ETH, Zurich. He holds the Lead Guest Chair Professorship of the Group on Wireless Communication and Networking at Tsinghua University, Beijing. He is an Honorary Professor at IIT Hyderabad.